\documentclass[letterpaper,prb,preprint]{revtex4-1}

 \usepackage{natbib}
 \usepackage{graphicx}
 \usepackage{dcolumn}
 \usepackage{amsmath}
  \usepackage{textcomp}
 \usepackage{float}
 \usepackage{rotating}
\begin{document}

\title{Influence Spatial Degeneracy on Rotational Spectroscopy Three Wave Mixing and enantiomeric state separation of Chiral Molecules }
 \author{Kevin K. Lehmann}
 \affiliation{
 Departments of Chemistry \& Physics, University of Virginia, Charlottesville VA, 22904-4319}
 \email{lehmann@virginia.edu}
\begin{abstract}
 \date{\today}
 Pulse flip angles are calculated for three wave mixing, three state cycles of chiral molecules 
 to produce optimized free induction decay amplitudes proportional to the enantiomeric excess of a sample, and
 to produced optimized degree of state specific enantiomeric separation.  These calculations account for the
 spatial degeneracy of the levels involved and the resulting inhomogeneous distribution of transition dipole moments.
 It is found that cycles of transitions that include R followed by Q followed by P branch transitions display only modest
 reductions of the calculated optimal signals if spatial degeneracy is ignored.  Transitions cycles P - Q - R are only
 slightly worse, while the Q - Q - Q cycles are much worse, increasingly so as the rotational total quantum number
 increases.
   \end{abstract}
 
 \maketitle

Molecular Chirality plays a major role in Chemistry and Biology and has been of interest since the pioneering work of Pasture.~\cite{Pasteur1864}
Traditional Spectroscopic methods of probing the chiral composition of mater has suffered from limited sensitivity due to the
fact that the most commonly used phenomena, Optical Rotation and Circular Birefringence, arise from interference of the
transition electric and magnetic dipole moments.~\cite{KauzmannQM}   It has long been known that, uniquely for bulk liquid and gases, Chiral
Molecules can generate three wave mixing signals and the amplitude of the generated waves are proportional to the 
enantiomeric excess of the sample.~\cite{Giordmaine65}  However, it is not possible to phase match this three wave mixing and thus these
signals have been weak as well.~\cite{Fisher00}

Starting with a proposal by Hirota,~\cite{Hirota12} followed by an experimental realization by Patterson and Doyle,~\cite{Patterson13b} microwave wave 
three wave mixing has recently been developed, which can be used to quantitatively measure enantiomeric excess of a particular 
Chiral molecular conformer.~\cite{Patterson13,Patterson13b,Patterson14,Shubert14a,Shubert14b,Lobsiger15,Schubert15,Schmitz15,Shubert16}   More recently, a small state dependent enantiomeric excess was generated
using a similar technique.~\cite{Eibenberger17,Perez17}  Much of the physical basis of these experiments can be understood using models of 
based upon transitions among three levels.~\cite{Patterson13b,Lobsiger15,Pratt17} However, such models ignore the complication caused by the $2J+1$ spatial degeneracy 
of molecules in free space with total angular momentum quantum number $J$.  

It is the purpose of this paper to present the results of calculations that predict the optimal excitation conditions
that produce signals for both enantiomeric excess determination and enantiomeric partial separation of an enantiomeric 
pair into distinct rotational levels.  The results demonstrate that the spatial degeneracy leads to a reduction of signal
strengths, but the size of the reductions are modest for cycles of transitions include states that differ by $J$, but
are much more dramatic if the states involved are all have the same $J$ value.

The following section reviews the cases where spatial degeneracy is ignored.  The next section reviews the relevant 
properties of the rotational states of asymmetric top molecules.   That will be followed by a section that presents the
predicted sample three wave mixing and the enantiomeric separation as a function of the quantum states involved
in the transitions used.

\section{Review of Three State Results}

Consider three nondegenerate states of a molecule which are labeled $a$, $b$, and $c$ with electric dipole
transitions between each of them.  This requires a molecule without parity symmetry and thus the transitions must
be of a chiral system.~\cite{Hirota12}     By choice of the phase of the wavefunctions, we can
select two of the transition dipole matrix elements to be positive, say $\mu_{ab}$ and $\mu_{bc}$.   The sign of
the third matrix element, $\mu_{ac}$ is then determined and will be opposite for an enantiomeric molecular pair.
We will label the enantiomers $R$ and $S$ depending upon the sign of the product of the three dipole.~\cite{Hirota12} 

The angular frequencies of the three transitions will be denoted as $\omega_{ab}, \omega_{bc}$, and $\omega_{ac}$ 
respectively.  Assume
that we can apply the rotating wave approximation and that the three transitions are sufficiently separated that we can
neglect off-resonance pumping.  Also assume that the transitions are driven by square pulses of lengths $\Delta t$
and with Rabi frequencies $\Omega_{ab} = E(\omega_{ab}) \mu_{ab} / 2 \hbar$, etc. where $E(\omega_{ab})$ is the
amplitude of the electric field oscillating at angular frequency $\omega_{ab}$.  We will denote the ``flip angle" of
a resonant pulse by $\theta_{ab} = \Omega_{ab} \Delta t$, but this is really a stand-in for $\int \Omega_{ab}(t) dt$.~\cite{Yariv}  

First, consider the case where all the population is initially in the state $a$ and neglect relaxation.   Excitation on resonance with the
$a-b$ transition with $\theta_{ab}$ will move half the population from $a \rightarrow b$
and  will produce maximum coherence between these states $\rho_{ab} = 0.5$, maximum oscillating
dipole moment, and free induction decay (FID) amplitude.~\cite{Yariv}   Follow this by radiation resonant on the $b-c$ transition with $\theta_{bc} = \pi$.
In addition to inverting the populations of the states $b$, and $c$, this pulse will quantitatively transfer the coherence $\rho_{ab}$
into a coherence $\rho_{ac}$, and this will result in a maximum strength FID at the $a-c$ resonant frequency.  This is a three wave mixing (3WM) signal.
The amplitude of this FID will be proportional
to $\mu_{ac}$ and thus will be exactly opposite in sign (or equivalently phase shifted by $\pi$) for the $R$ and $S$ enantiomers 
excited at the same place.  Thus, if we have a racemic sample (equal density of $R$ and $S$ forms), we have no net
 3WM FID signal.~\cite{Fisher00}  More generally, the amplitude of the FID will be proportional to the enantiomeric excess ($ee$)
which is defined as the difference in number of $R$ and $S$ molecules divided by the total number.    If we relax the assumption that
all the population was initially in the state $a$, we find that the FID is proportional to the initial value of $\rho^e_{aa} - \rho^e_{bb}$, i.e. the difference in
starting population between the states of the initially driven transition.

Following the $\pi$ pulse, the populations of the states $a$ and $c$ are equal.  If  the system is then driven with radiation resonant with
the $a-c$ transition, with the phase of the driving radiation such that it is out of phase with the $FID$ of the $R$ form, we can increase
the fraction of molecules in the $c$ state, while simultaneously decreasing the $c$ population for $S$ molecules.   If this third pulse
has flip angle of $\pi/2$ and the system started with all the population in state $a$, the system will end with all $R$ molecules in the $c$ state and all $S$ molecules
in the $a$ state.  This will be opposite if the phase of any of the pulses is shifted by $\pi$.  
This can be viewed as a form of chiral separation, though the separation is in the Hilbert Space of the quantum levels
instead of physical space.  Such a separation will be destroyed by single inelastic collisions.  In principle this separation can be
converted into an enrichment if we can selectively photochemically dissociate molecules from either the initial or final state,
$i.e.$ $a$ or $c$.  If we consider initial population in all three states, we find that this pulse sequence will cycle the initial populations
in enantiomer $R$   $\rho_{aa} \rightarrow \rho^e_{bb} , \rho_{bb} \rightarrow \rho^e_{cc}, \rho_{cc} \rightarrow \rho^e_{aa}$, while for enantiomer $S$
we will have inverted the initial populations in states $b$ and $c$, i.e. $\rho_{bb} \rightarrow \rho^e_{cc}, \rho_{cc} \rightarrow \rho^e_{bb}$. 
This produces an enantiomeric excess of $ee = \pm (\rho^{e}_{aa} - \rho^e_{bb} )(\rho^{e}_{aa} + \rho^e_{bb})$ in states $c$ and $a$
respectively.  A phase shift of $\pi$ for $E_{ac}$
will switch the signs of the enantiomeric excesses.

An alternative is to first drive the two-photon transition $a \rightarrow b \rightarrow c$.   If both $E_{ab}$ and $E_{bc}$ are simultaneously on
and constant amplitude for a time $\Delta t$ such that $\sqrt{\Omega_{ab}^2 +  \Omega_{bc}^2} \Delta t = \pi$, the population in state $b$
will end at the same value it started with.   If we also take $\Omega_{bc} = (\sqrt{2} \pm 1) \Omega_{ab}$, the two photon excitation will
generate the maximum $a-c$ coherence, $\rho_{ac} \rightarrow 0.5 \left( \rho^e_{aa} - \rho^e_{cc} \right)$ and equate the population in these levels, 
$\rho_{aa}, \rho_{cc} \rightarrow (1/2)\left( \rho^e_{aa} + \rho^e_{cc} \right)$.  
If the two photon transition is followed by a $\pi/2$ pulse on the $a - c$ transition, one can selectively de-excite one enantiomer 
(leaving all populations at their starting values) while inverting the initial populations of states $a$ and $c$ for the other enantiomer, leaving
state $b$ population unchanged.   Starting with a racemic mixture, this pulse sequence will produce an enantiomeric excess of 
$ee = \pm (\rho^{e}_{aa} - \rho^e_{cc} )(\rho^{e}_{aa} + \rho^e_{cc})$ in states $c$ and $a$ respectively.   
	
Another approach was suggested by Li and Bruder,\cite{Li08} who proposed excitation of the system with a $\pi/2$ pulse at $\omega_{ac}$,
followed by simultaneous excitation at $\omega_{ab}$ and $\omega_{bc}$ (with equal Rabi frequencies), followed by $\pi/2$
excitation again at $\omega_{ac}$.  This will drive $\rho_{aa} \rightarrow \rho_{bb}$ for one enantiomer and $\rho_{aa} \rightarrow
\rho_{cc}$ for the other.  The first two pulses is the time reverse of the two photon case considered in the previous paragraph, while
the last pulse move the initial $S$ population from state $a \rightarrow b$.  

\section{Review of Asymmetric Top Transitions}
We will now specialize to the case that the three states involved in the cycle are rotational levels of the ground vibronic state of a 
molecule.~\cite{Townes55}   Based upon the selection rules to be discussed below, a molecule can have a closed cycle of three dipole allowed rotational transitions
only if the molecule has nonzero projections of its electric dipole  on all three inertial axes.   We labels these projections $\mu_a, \mu_b, \mu_c$ for the projections onto
the inertial axes $A, B, C$ respectively.  A molecule that satisfies this condition must lack any point group
symmetry elements and belong to the point group $C_1$.  Such molecules are Chiral and also asymmetric tops.
The inertial tensor is second rank, and as a result we are free to switch the directions of the axes.  We can pick two to
have positive dipole projections (say $\mu_a$ and $\mu_b$) but the direction of the third axes is fixed by the condition that
the $A,B,C$ axes form a right handed system.  If we consider an enantiomeric pair of molecules, the product of the
three dipole projections will be positive for one, and negative for the other.  

The rotational eigenstates of an asymmetric top can be labeled with three quantum numbers: $J$, which is total angular momentum
quantum number, $\tau = 0,\ldots 2J+1$, and $M$ which is the quantum number for projection of the total angular momentum on
to some laboratory fixed axis.  For fixed $J$, the states are ordered the same as $\tau$, i.e. $E(J,\tau, M) < E(J, \tau', M)$ if
$\tau < \tau'$.  States with different values of $\tau$ are never exactly degenerate, though most states occur in pairs with
spacings that are much less than the average spacing.  These are known as asymmetry doublets and the splitting arises from
tunneling.    The states with fixed $J, \tau$ are $2J+1$ degenerate (different values of $M$) in isotropic space and form one
of the irreducible representations of the orthogonal group, $O(3)$.

For purposes of assigning symmetry labels to the states, it is convenient to introduce two additional labels
$K_p$ = Floor$[ (\tau +1)/2]$ and $K_o = $Floor$[ (2J + 1 - \tau)/2]$.  $K_p$ and $K_o$ are also commonly written as $K_a$ and
$K_c$ respectively.  The values of $K_p$ run $0, 1, 1, 2, 2,\ldots J, J$ and $K_o = J, J, J-1, J-1,\ldots 0,0$ as $\tau$ runs $0$ to $2J$.
The sum $K_p + K_o$ equals either $J$ or $J+1$ for all states. 
The symmetry of states can be assigned to four irreducible representations depending upon the whether $K_p$ and $K_o$ are
even or odd integers, $i.e.$ $(K_p,K_o)$ being (even,even), (even, odd), (odd, even) or (odd, odd).   If we have states $J,K_p,K_0,M$ and
$J',K'_p,K'_o,M'$, a transition between them is only allowed  if $J - J' = 0, \pm 1$ (but not if $J = J' = 0$); $M-M' = 0, \pm 1$; and if ($K_p - K'_p, K_o - K'_o)$
is (even,odd), in which case the transition dipole matrix element is proportional to $\mu_a$, is (odd, even), in which case the transition dipole
element is proportional to $\mu_c$, or (odd, odd), in which case the transition dipole element is proportional to $\mu_b$.   

Given these selection rules, three states $(J_1, \tau_1, M_1), (J_2, \tau_2, M_{2} )$, and $(J_3,\tau_3, M_3)$  can make up a closed cycle 
of allowed transitions only if the following conditions can be met, possibly with permutation of the label):  
1) $J_1 = J_2 = J_3$ (which we will label a QQQ case), $J_2 = J_3 = J_1+1$
(which we call a RQP case) or $J_2 = J_3 = J_1 -1$ (which we call a PQR case).
2) The three transitions $1 \leftrightarrow 2, 2 \leftrightarrow 3$, and $1 \leftrightarrow 3$ are each allowed by a distinct dipole projection,
i.e. are polarized by some permutation of molecular fixed axes $A, B, C$.  3) The electric fields driving the transitions must all be 
parallel or all perpendicular to each other.   If the sample starts with zero orientation in the states, i.e. the population of states that
different only in the sign of $M$ are equal, the case where all three electric fields are parallel will produce zero net three wave mixing
signals nor allow for enantiomeric separation, so for the rest of this paper, we will only consider the case of excitation (or detection)
of three orthogonally polarized waves.

The eigenfunctions of an asymmetric top can be expanded in Symmetric Top eigenfunctions.   
\begin{equation}
| J, \tau, M> = \sum_K  A_{J \tau K} | J, K, M>
\end{equation}
Note that the amplitudes, $A_{JK}$, do not depend upon $M$ and can be selected to be real.  The matrix elements of the electric dipole interaction when
driven by a resonant electric field $\vec{E}$ is
\begin{eqnarray}
\left< J, \tau, M | - \vec{E} \cdot \vec{\mu}_e | J', \tau', M' \right>  &=& - \left(  \sum_{G = X,Y,Z} E_G \phi_{G}(J M, J'M') \right) \times \nonumber \\
 &&  \left(  \sum_{g= a,b,c} \mu_g  \sum_{K,K'} A_{J \tau K} A_{J' \tau' K'} \, \phi_g(J K, J' K') \right) 
\end{eqnarray}
We will take quantization axes along $Z$ in the laboratory and along $A$ in the molecular frame, though all observables 
are independent of how axes are assigned (as long as one continues to use right handed systems in both frames).  
In that case, the nonzero direction cosine matrix elements are given in Table \ref{direction_cosine}.  Note, the sums over $K, K'$ will be
nonzero for at most one value of $g$ for a given pair of asymmetric top levels.  
For future notational convenience we will introduce notation
\begin{equation}
\left< J, \tau || \phi || J', \tau' \right> =  \sum_{K,K',g} A(J,K) A(J', K') \phi_g(J K, J' K')  \label{reducedphi}
\end{equation}

\section{Three wave mixing with spatial degeneracy}

Consider a triplet of levels $(J_1, \tau_1), (J_2, \tau_2)$ and $(J_3, \tau_3)$ driven on resonance with mutually perpendicular 
resonant fields of amplitudes $E_{12}, E_{23}$, and $E_{13}$.  We define the polarization directions as
$X$ for $E_{12}$, $Z$ for $E_{23}$ and $Y$ for $E_{13}$.  Define effective Rabi frequencies for each transition by
\begin{equation}
\Omega_{i,j} = E_{ij} \left< J_i, \tau_i || \phi || J_j, \tau_j \right> / 2 \hbar  \label{Omegaij}
\end{equation}
We can determine the time evolution for any period of constant field amplitudes by solving for the eigenvalues of 
the dress state Hamiltonian.~\cite{Shirley65}  This is in a $2(J_1+J_2 + J_3) + 3 $ dimensional space with matrix elements:

\begin{eqnarray}
H(J_1 M_1, J_2 M_2)  &=& \Omega_{12} \, \phi_X(J_1 M_1, J_2, M_2)  \nonumber \\
H(J_2 M_2, J_3 M_3)  &=& \Omega_{23} \,  \phi_Z(J_2 M_2, J_3, M_3) \nonumber \\
H(J_1 M_1, J_3 M_3)  &=& \Omega_{23} \,  \phi_Y(J_1 M_1, J_3, M_3)
\end{eqnarray}

This allows us to define $H$ as a function of the three Rabi frequencies, $\Omega_{12}, \Omega_{23}, \Omega_{13}$.
Define $E(\Omega_{12}, \Omega_{23}, \Omega_{13})$ as the Eigenvalues of $H$ and $B(\Omega_{12}, \Omega_{23}, \Omega_{13})$
the matrix with normalized eigenvectors.   For a period of duration $\Delta t$ with constant driving amplitudes, the 
time evolution operator, $U(\Omega_{12}, \Omega_{23}, \Omega_{13}, \Delta t)$ has matrix elements
\begin{equation}
U_{J_i M_i, J_j M_j} = \sum_{J_k M_k} B_{J_i M_i, k} \bar{B}_{J_j, M_j, k} \exp \left( - i E_k \Delta t \right)
\end{equation}
This will transform the density matrix $\rho \rightarrow \bar{U} \rho U$.  The electric dipole moment oscillating at
the resonance frequency of transition $i, j$ is given by
\begin{equation}
\mu_{i,j} = \sum_{M_i, M_j} \rho_{J_i M_i, J_j M_j} \phi_G (J_i M_i, J_j, M_j)   \label{muij}
\end{equation}
where $G = X, Z, Y$ for $(i,j) = (1,2), (2,3)$, and $(1,3)$.  
We also define population in level $i$ by $P_i = \sum_M \rho_{J_i M, J_i M}$.

We first treat the $RQP$ cases, with $J_2 = J_3 = J_1 + 1$, and where the entire population is distributed equal among
the $M$ levels of level 1, and level 2 and 3 initially unoccupied. We first consider excitation with only $E_{12}$ and optimize $\Omega_{12} \Delta t$ to produce the 
maximum absolute value of  $\mu_{12}$.  This is the equivalent of the $\pi/2$ pulse of the nondegenerate case.   We then apply radiation at
$E_{12}$ with $\Omega_{12} \Delta t$ adjusted to optimize the absolute magnitude of $\mu_{13}$.  Because of the sign change of the product
$\mu_a \mu_b \mu_c$ between enantiomers, the resulting value of $\mu_{13}$ will be opposite in sign for an enantiomer pair.  Table \ref{RQP} shows
the optimal flip angles and the resulting $\mu_{13}$ values in units of $\mu_g \left< J_1, \tau_1 || \phi_g || J_3, \tau_3 \right> P_1$ for $J_1 = 0 - 9$.  Simultaneous 
optimization of $\Omega_{12} \Delta t$  and $\Omega_{23} \Delta t$ to optimize $\mu_{13}$ changed the optimal value of
$\mu_{13}$ by less than $0.1\%$ in all cases.   After the application of the $E_{12}$ and $E_{23}$ pulses, a pulse at
$E_{13}$ was applied.   The value of $\Omega_{13} \Delta t$ of this pulse was optimized to produce the largest calculated population difference in level 3 
for application of field $\Omega_{23}$ compared to application of $-\Omega_{23}$ which corresponds to the degree of enantiomeric population
seperation.   

It can be seen that the case of $J_1 = 0$ produces ``ideal" results, i.e. the same as the nondegenerate cases discussed earlier.   Here, the system is
always in a pure state and pulse amplitudes can be adjusted to produce the desired result.    For $J_1 > 0$, we have a mixed state case and the spread
of Rabi frequencies as a function of $M$ results in some decrease in the size of the optimized results.  The decreases 
are modest.  For an enantiomerically pure sample, the 3-wave mixing polarization on the $1-3$ transition, as a function of $J_1$ rapidly approaches a value of $\sim 64\%$ of
what would be expected for driving the $1-3$ transition directly with a pulse of optimal area.
For the enantiomeric enrichment of a racemic sample,
the optimized enantiomeric excess in the level 3 falls from $1$ for $J_1 =1$ to about $0.5$ in the high $J_1$ limit.  The optimal value
of $\Omega_{13} \Delta t$ for the second $E_{13}$ pulse is almost identical that that of the first $E_{ac}$. While this expected for the pure
state case, it was not obvious that this would be the case for the cases with spatial degeneracy.   

These calculations overestimate the expected effects for two reasons.   The most important, in practice, is the inability to phase match the 3WM signals.  
The FID produced by the two pulses and the population modulation produced by the three pulses have a spatial dependence, $\vec{R}$, of
$\exp \left( i  \Delta \vec{k} \cdot \vec{R} \right)$.  For the case where $E_1 <  E_3 < E_2$, $\Delta \vec{k} = \vec{k}_{12} - \vec{k}_{23} - \vec{k}_{31}$
where $\vec{k}_{ij}$ is the wave vector for the radiation driving the $i \leftrightarrow j$ transition.  
More generally, we have a positive sign for $\vec{k}_{ij}$ if $E_i < E_j$ and
a negative sign in the opposite case.  If we could make the $\vec{k}$ all parallel, $\Delta \vec{k}$ would be zero and the signals produced at each point
in space would constructively interfere.   However, because we must use three perpendicularly polarized radiation fields, and $\vec{k}$ must be
perpendicular to $\vec{E}$ for each field, the three wavevectors cannot be
parallel -- at least one must be perpendicular to the other two.  We can minimize the dephasing caused by this lack of phase matching 
by taking energy levels where one of the spacings is as
small as possible, which practically means that transition is across an asymmetry doublet.  Assume that the $2-3$ transition is the lower frequency one,
 and that we take $\vec{k}_{12}$
 and $\vec{k}_{31}$ to be parallel and $\vec{k}_{23}$ to be perpendicular to these.  
Then $| \Delta \vec{k} | = \sqrt{2} | \vec{k}_{23}| = \sqrt{2} \omega_{13} / c$.
We can also minimize $| \Delta \vec{k} | $ by taking $\omega_{13}$ small, however, since the $1-3$ FID field amplitude produced by a fixed polarization is proportional to
$\omega_{13}$, this will likely will not be optimal.  

Another consideration is that chiral molecules have sufficient number of low energy rotational states
that the assumption that only the initial level of the transitions is populated is unlikely to be even approximately true, even at rotational temperatures of $\sim 1^{\circ}$\,K that
can be realized in pulsed supersonic expansions.  Population initially in state $2$ will produce opposite polarizations and enantiomeric enrichment.  If the
level populations can be described by a rotational temperature $T_r$, then the polarization produced on the $1-3$ will be reduced by a factor of 
$\left[1 - \exp( - \hbar \omega_{12} / k_b T_r) \right]$.  Of course the $1-3$ FID produced by direct excitation of that transition will also be reduced by a 
factor of $\left[1 - \exp( - \hbar \omega_{13} / k_b T_r) \right]$, which will be almost the same if we assume that $\omega_{23} << \omega_{12}, \omega_{13}$.
The degree of enantiomeric separation, will be reduced by a factor of 
$\tanh ( - \hbar \omega_{12} / 2 k_b T_r) $ for state $1$.

In practical experiments, one can arrange the apparatus (say by using dual polarization horns) to be able to observe the $1-2$ and $1-3$ FID signals and thus to 
determine the pulse lengths that optimize these signals.  However, it will often be difficult to observe the $2-3$ FID as this is at a low frequency and also is radiating
in a direction perpendicular to the other two transitions.  The $2-3$ excitation flip angle can be calibrated by determining the $2-3$ pulse that nulls the $1-2$ 
FID produced by an optimized pulse area.  For the pure state $J_1 = 0$ case, this is equal to to the pulse that maximizes the $1-3$ polarization, but this is not
true for higher values of $J_1$.  Table \ref{RQP} lists the values of $\Omega_{23} \Delta t$ that nulls the $1-2$ FID.  This pulse produces a $1-3$ polarization 
about $\sim 10\%$ lower than the optimal $\Omega_{23}$, but one can the tabulated values to scale the $\Omega_{23}$ value from that which
cancels the $1-2$ FID to one close to optimal.   

It will generally be easier to realize the $\omega_{23} << \omega_{12}, \omega_{13}$ condition if we take a $RQP$ cycle, but there are also cases (such nearly
prolate tops), where this can also be realized for a $QQQ$ cycle.  Table \ref{QQQ} gives the results, similar to those of Table \ref{RQP}, but for such a cycle.  Obviously, the 
minimum $J_1$ value for such a cycle is $J_1 = 1$.  Comparison of the two tables shows that the FID and enantiomeric separations are predicted to be quite a bit smaller for 
$QQQ$ cycles compared to $RQP$ cycles.  The ratio of the predicted $\omega_{13}$ for the QQQ cases divided by those for the RQP cases (other
factors being constant) fall from 0.44 for $J_a =1$ to 0.06 for $J_a = 10$.   If we take the $Z$ axis parallel to $E_{23}$, then we can see that in the $RQP$ case, all three transitions are
favored for high $M$ values, but in the $QQQ$ cases, two of the transitions are favored for low $|M|$ values and one for high $|M|$ values.  This enhances
the degree that $1-3$ polarization varies as a function of $M$.   Likewise, the optimal values for the degree of enantiomeric separation, as
measured by $\Delta \rho_{33}$, is also substantially less for the QQQ cases compared to the RQP cases.   These results suggest that use of a QQQ cycle is highly unlikely to be optimal.  
An interesting result is that the value of the $\Omega_{23} \Delta t$ that cancels the $1-2$ FID is predicted to be about twice as large as the one that produces the
maximum $1-3$ polarization, and that this pulse also gives a null for that $1-3$ polarization.   The fractional impact of finite temperature is the same as for the $RQP$ cases.  Table \ref{PQR}
shows the results for PQR cycles.   These are are similar to but slightly less favorable than the $RQP$ cases.

As is the case for the ideal, 3-state model, we can also create the $1-3$ polarization by a two photon excitation, i.e. with both $E_{12}$ and $E_{23}$
simultaneously present.  The values of $\Omega_{12} \Delta t$ and $\Omega_{23} \Delta t$ were optimized to produce the largest 
calculated values of $\mu_{13}$. and these are presented in Table \ref{RQP2} along with the resulting $1-3$ polarization.    Comparison of Tables \ref{RQP} \& \ref{RQP2} shows that there is
little difference in the size of the optimized polarization between sequential and simultaneous excitation of the $1-2$ and $2-3$ transition.  
The same folds to the optimized enantiomeric excitation predicted by a following $1-3$ pulse.  It is less clear how to experimentally optimize the values
of $E_{12}$ and $E_{23}$ in this case if the resulting $1-3$ FID is weak, say because the sample has a low initial $ee$ value.  If we take $\Delta t$ as equal
for the two sequential pulses and $\Delta t$ twice as long for the simultaneous excitation, i.e. the same total time, then the optimal field amplitudes required 
for the simultaneous case is modestly lower than for the sequential case.

 Table \ref{QQQ2} gives the optimized results for simultaneous excitation in the QQQ cases.  As for sequential excitation, the 3WM FID and enantiomeric separation 
 predicted for the QQQ cases is far less than for the $RQP$ cases, and so these are unlikely to be optimal in experimental practice.  Table \ref{PQR2} shows the 
 results for two photon excitation for these cycles.  Like for the two pulse excitation, the PQR cycles gives FID and enantiomeric specific population transfers that
 are modesty lower than those of the RQP cycles. 
 
 \section{Conclusions}
 
 Three wave mixing of molecular rotational transitions of chiral molecules has been demonstrated as a method to measure the enantiomeric excess of a sample.  More recently, three wave excitation has
 been demonstrated to create an enantiomeric excess in particular rotational levels of chiral molecules.   Qualitatively, these experiments can be understood using a model of a  3-state cycle of transitions, but such a
 model ignores the unavoidable spatial degeneracy of rotational states in isotropic space.   The present work has presented the results of calculations that determine optimal 
 excitation conditions and the resulting signals including the full set of states involved in a cycle of transitions between three rotational levels.   The results establish that in all cases, except those involving the non-degenerate $J=0$ state, there is a reduction in the size of the signals and state dependent enantiomeric excess compared to
 what is predicted by the simple 3-state model.   However, the reduction is modest for cycles that involve two $\Delta J = \pm 1$ transitions.  The reductions are much larger and
 appear to tend to zero in the high $J$ limit for cycles that involve three levels all with the same total angular momentum quantum number.   It is expected that the
 present results will be useful to those designing and optimizing such experiments.
 
\section{Acknowledgments}
The author many insightful discussions with Brooks Pate.  This research was supported by NSF grant CHM 1531913.

\bibliography{Chirality}

\begin{table}[htp]
\caption{Direction Cosine Matrix Element Factors\cite{Townes55}}
\begin{center}
\begin{tabular}{|c|c|c|c| c|c|}
\hline
 &  J' = J-1  &  J' = J   & J' = J+1\\
\hline
$\phi_a(J,M,J', M)    $ &$   \frac{ \sqrt{J^2 - M^2}}{ (J^2 \left[ 4J^2 - 1)\right]^{1/4}}$ & $\frac{M}{\sqrt{J(J+1) } }$   &  $\frac{ \sqrt{(J+1)^2 - M^2}}{ \left[(J+1)^2 (2J+1)(2J+3)\right]^{1/4}}$  \\ \hline
$\phi_b(J,M, J', M \pm 1)$ &$ \pm  \frac{(J \mp M)(J \mp M-1)}{2 \left[ J^2 ( 4J^2 - 1)\right]^{1/4}}   $  &  $ \frac{ \sqrt{(J \mp M)(J \pm M+1)}}{2 \sqrt{J(J+1) }}  $ &$ \mp \frac{ \sqrt{(J \pm M+1)(J \pm M+2)  } }{ 2\left[(J+1)^2 (2J+1)(2J+3)\right]^{1/4} } $\\  \hline
$\phi_c(J,M, J',M \pm 1) $&$ -i \frac{(J \mp M)(J \mp M-1)}{2 \left[ J^2 ( 4J^2 - 1)\right]^{1/4}} $  &$  \mp i \frac{ \sqrt{(J \mp M)(J \pm M+1)   }}{2 \sqrt{J(J+1) }}  $ &$ i \frac{ \sqrt{(J \pm M+1)(J \pm M+2) }}{ 2\left[(J+1)^2 (2J+1)(2J+3)\right]^{1/4} } $\\  \hline
\multicolumn{4}{l}{$ \phi_Z(J,M,J', M) = \phi_a(J,M,J', M), \,\, \phi_X(J,M,J', M) = \phi_a(J,M,J', M), \,\, \phi_Y(J,M,J', M) = \phi_c(J,M,J', M)$}

\end{tabular}
\end{center}
\label{direction_cosine}
\end{table}

\begin{sidewaystable}[htp]

\begin{center}
\begin{tabular}{|c|c|c|c|c|c|c|c|c|c|c|c|c|c|c|c|}
\hline																						
\multicolumn{2}{|c|}{}&		\multicolumn{2}{c|}{ $<\mu_{12}>$ Optimized}		&	\multicolumn{4}{c|}{ $<\mu_{13}>$ optimized}		&			\multicolumn{4}{c|}{$\Delta P_3$ optimized}	&			\multicolumn{3}{c|}{$<\mu_{12}>$ cancelled}	  
		\\	\hline
$J_1$ &	$J_2 = J_3$  &	   $\Omega_{12} \Delta t$ &	$<\mu_{12}>$	&	$\Omega_{23} \Delta t$ &	$<\mu_{12}>$	& $<\mu_{13}>$ &	$<\mu_{23}>$	&  $\Omega_{13} \Delta t$ & $ P_3$(R)	&	$P_3$(S)	& $\Delta P_3$   &	$\Omega_{23} \Delta t$	&  $<\mu_{13}>$ &	$<\mu_{23}>$	\\ \hline
0&1&1.0336&0.3799&2.2214&-0.3695&0.3799&0.0000&1.0336&1.0000&0.0000&1.0000&2.2214&0.3799&0.0 \\
1&2&1.1886&0.3252&2.1365&-0.2937&0.2685&0.0103&1.1922&0.8668&0.0527&0.8141&2.4802&0.2604&0.0667  \\
2&3&1.2234&0.3063&2.1310&-0.2456&0.2372&0.0069&1.2307&0.8239&0.0828&0.7412&2.5872&0.2248&0.0647 \\
3&4&1.2364&0.2965&2.1299&-0.2235&0.2219&0.0051&1.2457&0.7980&0.0967&0.7013&2.6438&0.2074&0.0618  \\
4&5&1.2425&0.2904&2.1295&-0.2117&0.2128&0.0040&1.2533&0.7808&0.1045&0.6764&2.6782&0.1970&0.0597  \\
5&6&1.2459&0.2863&2.1293&-0.2046&0.2068&0.0032&1.2578&0.7688&0.1094&0.6593&2.7013&0.1902&0.0582  \\
6&7&1.2479&0.2833&2.1292&-0.2000&0.2025&0.0027&1.2607&0.7599&0.1129&0.6470&2.7177&0.1853&0.0571  \\
7&8&1.2492&0.2810&2.1291&-0.1967&0.1993&0.0023&1.2627&0.7531&0.1154&0.6377&2.7300&0.1817&0.0562  \\
8&9&1.2502&0.2793&2.1290&-0.1943&0.1968&0.0020&1.2641&0.7477&0.1173&0.6303&2.7395&0.1789&0.0556  \\
9&10&1.2508&0.2778&2.1290&-0.1925&0.1948&0.0018&1.2652&0.7433&0.1189&0.6244&2.7471&0.1767&0.0551  \\
10&11&1.2513&0.2767&2.1289&-0.1911&0.1931&0.0016&1.2661&0.7397&0.1201&0.6196&2.7533&0.1749&0.0546  \\ \hline 
\end{tabular}
\end{center}
\caption{Chiral enrichment by $\pi/2$ pulse on 1-2 transition, followed by $\pi$ pulse on 2-3 transition, followed by $\pi/2$ pulse on 1-3 transition: RQP cases.  $\Omega_{ij}$ are defined by Eq. \ref{Omegaij},
$\Delta t$ the length of the excitation pulse, $\mu_{ij}$ by Eq. \ref{muij}, and $P_i$ the fractional population in state $i$.  Calculations used initial conditions $P_i = \delta_{i,j}$}
\label{RQP}
\end{sidewaystable}

\begin{sidewaystable}[htp]
\begin{center}
\begin{tabular}{|c|c|c|c|c|c|c|c|c|c|c|c|c|c|}
\hline																						
&		\multicolumn{2}{c|}{ $<\mu_{12}>$ Optimized}		&	\multicolumn{4}{c|}{ $<\mu_{13}>$ optimized}		&			\multicolumn{4}{c|}{$\Delta P_3$ optimized}	&			\multicolumn{3}{c|}{$<\mu_{12}>$ cancelled}	  
		\\	\hline
$J_1 = J_2 = J_3 $  &	   $\Omega_{12} \Delta t$ &	$<\mu_{12}>$	&	$\Omega_{23} \Delta t$ &	$<\mu_{12}>$	& $<\mu_{13}>$ &	$<\mu_{23}>$	&  $\Omega_{13} \Delta t$ & $ P_3$(R)	&	$P_3$(S)	& $\Delta P_3$   &	$\Omega_{23} \Delta t$	&  $<\mu_{13}>$ &	$<\mu_{23}>$	\\ \hline
 1&1.1107&0.2357&2.2214&-0.2357&0.1179&0.0000&1.1107&0.5000&0.1667&0.3333&4.4429&0.0000&0.0000 \\
2&1.0615&0.2234&1.9989&-0.1078&0.0578&-0.0276&0.9433&0.3710&0.2318&0.1392&4.0671&0.0000&-0.0037  \\ 
3&1.0509&0.2207&1.9557&-0.0949&0.0395&-0.0307&0.9053&0.3419&0.2509&0.0910&4.0191&0.0000&0.0038 \\
4&1.0468&0.2196&1.9395&-0.0904&0.0301&-0.0317&0.8909&0.3289&0.2605&0.0684&4.0031&0.0000&0.0054 \\
5&1.0448&0.2191&1.9316&-0.0882&0.0244&-0.0321&0.8840&0.3215&0.2665&0.0550&3.9958&0.0000&0.0059 \\
6&1.0436&0.2188&1.9272&-0.0870&0.0206&-0.0324&0.8801&0.3166&0.2705&0.0461&3.9917&0.0000&0.0062 \\
7&1.0429&0.2186&1.9244&-0.0862&0.0178&-0.0325&0.8776&0.3131&0.2734&0.0397&3.9893&0.0000&0.0064 \\
8&1.0424&0.2185&1.9226&-0.0857&0.0157&-0.0326&0.8760&0.3106&0.2757&0.0349&3.9877&0.0000&0.0065 \\
9&1.0421&0.2184&1.9213&-0.0854&0.0140&-0.0327&0.8749&0.3086&0.2774&0.0311&3.9865&0.0000&0.0065 \\
10&1.0419&0.2184&1.9204&-0.0851&0.0126&-0.0327&0.8741&0.3070&0.2789&0.0281&3.9857&0.0000&0.0066 \\
 \hline 
\end{tabular}
\end{center}
\caption{Chiral enrichment by $\pi/2$ pulse on 1-2 transition, followed by $\pi$ pulse on 2-3 transition, followed by $\pi/2$ pulse on 1-3 transition: QQQ cases.
$\Omega_{ij}$ are defined by Eq. \ref{Omegaij},
$\Delta t$ the length of the excitation pulse, $\mu_{ij}$ by Eq. \ref{muij}, and $P_i$ the fractional population in state $i$.  Calculations used initial conditions $P_i = \delta_{i,j}$	}
\label{QQQ}
\end{sidewaystable}
  
\begin{sidewaystable}[htp]
\begin{center}
\begin{tabular}{|c|c|c|c|c|c|c|c|c|c|c|c|c|c|c|c|}
\hline																						
\multicolumn{2}{|c|}{}&		\multicolumn{2}{c|}{ $<\mu_{12}>$ Optimized}		&	\multicolumn{4}{c|}{ $<\mu_{13}>$ optimized}		&			\multicolumn{4}{c|}{$\Delta P_3$ optimized}	&			\multicolumn{3}{c|}{$<\mu_{12}>$ cancelled}	  
		\\	\hline
$J_1$ &	$J_2 = J_3$  &	   $\Omega_{12} \Delta t$ &	$<\mu_{12}>$	&	$\Omega_{23} \Delta t$ &	$<\mu_{12}>$	& $<\mu_{13}>$ &	$<\mu_{23}>$	&  $\Omega_{13} \Delta t$ & $ P_3$(R)	&	$P_3$(S)	& $\Delta P_3$   &	$\Omega_{23} \Delta t$	&  $<\mu_{13}>$ &	$<\mu_{23}>$	\\ \hline
2&1&1.1886&0.1951&2.2214&-0.1817&-0.0976&0.0000&1.1947&0.1895&0.4860&-0.2965&2.9061&-0.0864&0.0596 \\
3&2&1.2234&0.2188&2.1480&-0.1770&-0.1210&-0.0036&1.2451&0.1729&0.5554&-0.3825&2.9485&-0.1023&0.0443\\
4&3&1.2364&0.2306&2.1363&-0.1745&-0.1342&-0.0036&1.2609&0.1631&0.5922&-0.4291&2.9347&-0.1137&0.0432\\
5&4&1.2425&0.2376&2.1326&-0.1735&-0.1424&-0.0032&1.2673&0.1570&0.6145&-0.4575&2.9184&-0.1214&0.0437\\
6&5&1.2459&0.2422&2.1310&-0.1733&-0.1480&-0.0029&1.2703&0.1529&0.6294&-0.4765&2.9048&-0.1268&0.0444\\
7&6&1.2479&0.2455&2.1302&-0.1734&-0.1521&-0.0026&1.2718&0.1499&0.6400&-0.4900&2.8939&-0.1309&0.0450\\
8&7&1.2492&0.2480&2.1298&-0.1737&-0.1551&-0.0023&1.2727&0.1477&0.6479&-0.5001&2.8851&-0.1340&0.0455\\
9&8&1.2502&0.2499&2.1295&-0.1739&-0.1575&-0.0021&1.2732&0.1460&0.6540&-0.5080&2.8780&-0.1364&0.0459\\
10&9&1.2508&0.2514&2.1293&-0.1742&-0.1594&-0.0020&1.2735&0.1446&0.6589&-0.5143&2.8721&-0.1384&0.0463\\
11&10&1.2513&0.2526&2.1292&-0.1745&-0.1610&-0.0018&1.2737&0.1435&0.6629&-0.5194&2.8672&-0.1400&0.0466 \\ \hline 
\end{tabular}
\end{center}
\caption{Chiral enrichment by $\pi/2$ pulse on 1-2 transition, followed by $\pi$ pulse on 2-3 transition, followed by $\pi/2$ pulse on 1-3 transition: PQR cases.
$\Omega_{ij}$ are defined by Eq. \ref{Omegaij},
$\Delta t$ the length of the excitation pulse, $\mu_{ij}$ by Eq. \ref{muij}, and $P_i$ the fractional population in state $i$.  Calculations used initial conditions $P_i = \delta_{i,j}$	}

\label{PQR}
\end{sidewaystable}

 \begin{sidewaystable}[htp]
\begin{center}
\begin{tabular}{|c|c|c|c|c|c|c|c|c|c|c|c|c|c|}
\hline																						
\multicolumn{2}{|c|}{}&	\multicolumn{8}{c|}{ $<\mu_{13}>$ Optimized by 2-photon trans.}   & \multicolumn{4}{c|}{$\Delta P_3$ optimized}	\\ \hline
$J_1$ & $J_2 = J_3$	&	   $\Omega_{12} \Delta t$ &	$\Omega_{23} \Delta t$ &	$<\mu_{12}>$	& $<\mu_{13}>$ &	$<\mu_{23}>$ 	& $P_1$ & $P_2$ &$P_3$ &
 $\Omega_{13} \Delta t$ & $ P_3$(R)	&	$P_3$(S)	& $\Delta P_3$   	\\ \hline
0&1&1.5822&4.1046&0.0000&0.3799&0.0000&0.5000&0.0000&0.5000& 1.0336&1.0000&0.0000&1.0000   \\
1&2&1.5938&3.6526&0.1044&0.2507&-0.0364&0.4897&0.1768&0.3335&1.1891&0.8113&0.0532&0.7581 \\
2&3&1.6669&3.6209&0.1033&0.2236&-0.0418&0.4816&0.1976&0.3208&1.2224&0.7805&0.0864&0.6941 \\
3&4&1.7171&3.6411&0.0989&0.2116&-0.0416&0.4768&0.2060&0.3172&1.2344&0.7630&0.1002&0.6629 \\
4&5&1.7505&3.6636&0.0953&0.2046&-0.0408&0.4743&0.2103&0.3154&1.2404&0.7516&0.1075&0.6440 \\
5&6&1.7738&3.6824&0.0926&0.2001&-0.0400&0.4729&0.2129&0.3143&1.2439&0.7435&0.1121&0.6313 \\
6&7&1.7907&3.6976&0.0905&0.1968&-0.0393&0.4721&0.2145&0.3134&1.2463&0.7375&0.1153&0.6222 \\
7&8&1.8034&3.7097&0.0889&0.1944&-0.0387&0.4716&0.2157&0.3127&1.2480&0.7328&0.1176&0.6152 \\
8&9&1.8133&3.7195&0.0877&0.1925&-0.0383&0.4714&0.2165&0.3121&1.2493&0.7291&0.1193&0.6098 \\
9&10&1.8212&3.7276&0.0867&0.1909&-0.0379&0.4712&0.2171&0.3117&1.2503&0.7260&0.1206&0.6054 \\
10&11&1.8276&3.7343&0.0859&0.1897&-0.0375&0.4712&0.2176&0.3112&1.2511&0.7235&0.1217&0.6018 \\
\hline 
\end{tabular}
\end{center}
caption{Chiral enrichment by two photon transition $1 \rightarrow 2 \rightarrow 3$ followed by excitation on the  $1 \rightarrow 3$ transition.: RQP cases.
$\Omega_{ij}$ are defined by Eq. \ref{Omegaij},
$\Delta t$ the length of the excitation pulse, $\mu_{ij}$ by Eq. \ref{muij}, and $P_i$ the fractional population in state $i$.  Calculations used initial conditions $P_i = \delta_{i,j}$	}

\label{RQP2}
\end{sidewaystable}

\begin{sidewaystable}[htp]
\begin{center}
\begin{tabular}{|c|c|c|c|c|c|c|c|c|c|c|c|c|}
\hline																						
&	\multicolumn{8}{c|}{ $<\mu_{13}>$ Optimized by 2-photon trans.}   & \multicolumn{4}{c|}{$\Delta P_3$ optimized}	\\ \hline
$J_1 = J_2 = J_3$	&	   $ \Omega_{12} \Delta t $ &	$ \Omega_{23} \Delta t $ &	$<\mu_{12}>$	& $<\mu_{13}>$ &	$<\mu_{23}>$ 	& $P_1$ & $P_2$ &$P_3$ &
 $\Omega_{13} \Delta t$ & $ P_3$(R)	&	$P_3$(S)	& $\Delta P_3$   	\\ \hline
1&1.7001&4.1046&0.0792&0.1179&0.0000&0.5433&0.2901&0.1667&1.1107&0.5000&0.1667&0.3333 \\
2&1.1313&3.1631&0.1603&0.0476&-0.0370&0.7127&0.1892&0.0981&0.9353&0.3275&0.2139&0.1137  \\
3&1.0585&2.9827&0.1640&0.0312&-0.0364&0.7380&0.1787&0.0833&0.9074&0.2949&0.2227&0.0721 \\
4&1.0336&2.9165&0.1649&0.0235&-0.0360&0.7468&0.1751&0.0782&0.8977&0.2815&0.2278&0.0536 \\
5&1.0219&2.8845&0.1653&0.0189&-0.0358&0.7509&0.1734&0.0758&0.8931&0.2741&0.2312&0.0429 \\
6&1.0155&2.8667&0.1655&0.0158&-0.0356&0.7531&0.1724&0.0744&0.8906&0.2695&0.2336&0.0359 \\
7&1.0115&2.8556&0.1656&0.0136&-0.0355&0.7545&0.1719&0.0736&0.8890&0.2663&0.2354&0.0309 \\
8&1.0089&2.8483&0.1657&0.0120&-0.0355&0.7554&0.1715&0.0731&0.8880&0.2640&0.2369&0.0271 \\
9&1.0071&2.8432&0.1657&0.0107&-0.0354&0.7561&0.1712&0.0727&0.8872&0.2622&0.2380&0.0242 \\
10&1.0058&2.8395&0.1657&0.0097&-0.0354&0.7565&0.1710&0.0724&0.8867&0.2608&0.2390&0.0218 \\
\hline 
\end{tabular}
\end{center}
\caption{Chiral enrichment by two photon transition $1 \rightarrow 2 \rightarrow 3$ followed by excitation on the  $1 \rightarrow 3$ transition.: QQQ cases.
$\Omega_{ij}$ are defined by Eq. \ref{Omegaij},
$\Delta t$ the length of the excitation pulse, $\mu_{ij}$ by Eq. \ref{muij}, and $P_i$ the fractional population in state $i$.  Calculations used initial conditions $P_i = \delta_{i,j}$	}
\label{QQQ2}
\end{sidewaystable}

\begin{sidewaystable}[htp]
\begin{center}
\begin{tabular}{|c|c|c|c|c|c|c|c|c|c|c|c|c|c|}
\hline																						
\multicolumn{2}{|c|}{}&	\multicolumn{8}{c|}{ $<\mu_{13}>$ Optimized by 2-photon trans.}   & \multicolumn{4}{c|}{$\Delta P_3$ optimized}	\\ \hline
$J_1$ & $J_2 = J_3$	&	   $\Omega_{12} \Delta t$&	$\Omega_{23} \Delta t$ &	$<\mu_{12}>$	& $<\mu_{13}>$ &	$<\mu_{23}>$ 	& $P_1$ & $P_2$ &$P_3$ &
 $\Omega_{13} \Delta t$ & $ P_3$(R)	&	$P_3$(S)	& $\Delta P_3$   	\\ \hline
2&1&1.8256&4.1100&0.0464&-0.0977&0.0013&0.6310&0.1646&0.2044&1.1947&0.1890&0.4858&-0.2968 \\
3&2&1.8210&3.8424&0.0733&-0.1211&-0.0260&0.5710&0.1905&0.2385&1.2449&0.1775&0.5601&-0.3826\\
4&3&1.8743&3.8562&0.0733&-0.1355&-0.0279&0.5359&0.2047&0.2594&1.2606&0.1651&0.5982&-0.4332\\
5&4&1.9000&3.8656&0.0727&-0.1443&-0.0286&0.5170&0.2113&0.2717&1.2652&0.1574&0.6202&-0.4628\\
6&5&1.9117&3.8679&0.0724&-0.1501&-0.0291&0.5060&0.2148&0.2793&1.2664&0.1525&0.6343&-0.4817\\
7&6&1.9167&3.8666&0.0725&-0.1541&-0.0295&0.4990&0.2168&0.2843&1.2664&0.1492&0.6440&-0.4948\\
8&7&1.9185&3.8638&0.0726&-0.1572&-0.0299&0.4943&0.2180&0.2878&1.2661&0.1469&0.6511&-0.5043\\
9&8&1.9187&3.8604&0.0729&-0.1595&-0.0302&0.4909&0.2188&0.2903&1.2657&0.1451&0.6566&-0.5115\\
10&9&1.9180&3.8571&0.0731&-0.1613&-0.0305&0.4885&0.2193&0.2922&1.2652&0.1437&0.6608&-0.5171\\
11&10&1.9170&3.8538&0.0734&-0.1627&-0.0308&0.4866&0.2197&0.2937&1.2648&0.1426&0.6643&-0.5217\\
\hline 
\end{tabular}
\end{center}
\caption{Chiral enrichment by two photon transition $1 \rightarrow 2 \rightarrow 3$ followed by excitation on the  $1 \rightarrow 3$ transition.: PQR cases.
$\Omega_{ij}$ are defined by Eq. \ref{Omegaij},
$\Delta t$ the length of the excitation pulse, $\mu_{ij}$ by Eq. \ref{muij}, and $P_i$ the fractional population in state $i$.  Calculations used initial conditions $P_i = \delta_{i,j}$	}
\label{PQR2}
\end{sidewaystable}

\end{document}